\documentclass{scrartcl}
\pdfoutput=1
\usepackage{amssymb,amsmath,amsthm, enumerate} 
\usepackage{graphicx}
\usepackage{algorithm}
\usepackage{algorithmicx}
\usepackage{algpseudocode}
\usepackage{subfigure}
\usepackage[numbers]{natbib}
\usepackage{fullpage}
\usepackage[nodisplayskipstretch]{setspace}
\usepackage[stable]{footmisc}
\usepackage[T1]{fontenc}
\usepackage[utf8]{inputenc}
\usepackage{authblk}
\usepackage{color}

\title{\textbf{Geodesic Lagrangian Monte Carlo over the space of positive definite matrices:}}
\subtitle{with application to Bayesian spectral density estimation}
\author[1]{Andrew Holbrook\thanks{aholbroo@uci.edu}}
\author[2]{Shiwei Lan}
\author[1]{Alexander Vandenberg-Rodes}
\author[1]{Babak Shahbaba}
\affil[1]{Department of Statistics\\ University of California, Irvine}
\affil[2]{Department of CMS\\ California Institute of Technology}

\begin{document}
\newcommand{\E}{\mathbb{E}}
\newcommand{\N}{\mathbb{N}}
\newcommand{\R}{\mathbb{R}}
\newcommand{\C}{\mathbb{C}}
\newcommand{\Z}{\mathbb{Z}}
\newcommand{\G}{\mathfrak{G}}
\newcommand{\m}{\mathcal }
\newcommand{\bb}{\mathbb }

\def\tens#1{\mathsf{#1}}

\def\v#1{\ensuremath{\mathchoice
                 {\mbox{\boldmath$\displaystyle\mathbf{#1}$}}
                 {\mbox{\boldmath$\textstyle\mathbf{#1}$}}
                 {\mbox{\boldmath$\scriptstyle\mathbf{#1}$}}
                 {\mbox{\boldmath$\scriptscriptstyle\mathbf{#1}$}}}}

\newcommand{\x}{{\v x}}
\newcommand{\y}{{\v y}}
\newcommand{\z}{{\v z}}

\renewcommand{\a}{\v a}
\renewcommand{\b}{\v b}
\renewcommand{\c}{\v c}
\renewcommand{\d}{\v d}
\newcommand{\Var}{\operatorname{Var}}
\newcommand{\Cov}{\operatorname{Cov}}
\newtheorem{theorem}{Theorem}
\newtheorem{prop}{Proposition}
\newtheorem{lemma}{Lemma}
\newtheorem{cor}{Corollary}
\newtheorem{defn}{Definition}
\newtheorem{remark}{Remark}
\newtheorem{example}{Example}
\newtheorem{question}{Question}
\newcommand{\vectwo}[2]{\begin{pmatrix}
#1 \\ #2
\end{pmatrix}}

\renewcommand{\thefootnote}{\fnsymbol{footnote}}
\renewcommand\rightmark{PDHMC}
\renewcommand\leftmark{Holbrook et al.}


\date{}
\maketitle

{\abstract We extend the application of Hamiltonian Monte Carlo to allow for sampling from probability distributions defined over symmetric or Hermitian positive definite matrices. To do so, we exploit the Riemannian structure induced by Cartan's century-old canonical metric.  The geodesics that correspond to this metric are available in closed-form and---within the context of Lagrangian Monte Carlo---provide a principled way to travel around the space of positive definite matrices.  Our method improves Bayesian inference on such matrices by allowing for a broad range of priors, so we are not limited to conjugate priors only. In the context of spectral density estimation, we use the (non-conjugate) complex reference prior as an example modeling option made available by the algorithm. Results based on simulated and real-world multivariate time series are presented in this context, and future directions are outlined.}


\section{Introduction} \label{intro}

 In statistics, positive definite (PD) matrices primarily appear as covariance matrices parameterizing the multivariate Gaussian models. This type of models is the workhorse of modern statistics and machine learning:  linear regression, probabilistic principal components analysis, Gaussian Markov random fields, spectral density estimation, and Gaussian process models all rely on the multivariate Gaussian distribution.  The $d$-dimensional Gaussian distribution is completely specified by a mean vector $\mu$ and a covariance matrix $\Sigma$ in $\mathcal{S}^+_d$, the space of $d$-by-$d$ PD matrices.  By imposing different structures on the covariance matrix, one can create different models.  In some cases, it is possible to parameterize the covariance matrices in terms of a small number of parameters. However, learning of the \emph{unstructured} covariance matrices, usually involved in inference on a large number of parameters, has remained as an issue. The conjugate Gaussian inverse-Wishart model has known deficiencies \cite{tokuda2011visualizing}.  Outside of non-linear parameterizations of the Cholesky decomposition or matrix logarithm, there has not yet been a way to perform Bayesian inference directly on the space of PD matrices with flexible prior specifications using unstructured covariance matrices.

In this most general context the difficulty is in sampling from a posterior distribution on an abstract, high-dimensional manifold with boundary.  It has not been clear how to propose moves from point to point within (and without leaving) this space. Our method takes advantage of the intrinsic, Riemannian geometry on the space of PD matrices.  This space is incomplete under the Euclidean metric: following a straight trajectory will often result in matrices that are no longer positive definite. The space is, however, geodesically complete under the canonical metric: no matter how far the sampler travels along any geodesic, it never leaves the space of PD matrices. Intuitively, we redefine `straight line' in a way that precludes leaving the set of interest.    Moreover, the metric-induced geodesics provide a natural way to traverse the space of PD matrices, and these geodesics fit nicely with recent advances in Hamiltonian Monte Carlo on manifolds \cite{girolami2011riemann,byrne2013geodesic,lan2014spherical}.

To this end, we use geodesic Lagrangian Monte Carlo (gLMC), which belongs to a growing class of Hamiltonian Monte Carlo algorithms. Hamiltonian Monte Carlo (HMC) provides an intelligent, partially deterministic method for moving around the parameter space while leaving the target distribution invariant.  New Markov states are generated by numerically integrating a Hamiltonian system while Metropolis-Hastings steps account for the numerical error \cite{neal2011mcmc}. Riemannian manifold Hamiltonian Monte Carlo (RMHMC) adapts the proposal path by incorporating second-order information in the form of a Riemannian metric tensor \cite{girolami2011riemann}.  Lagrangian Monte Carlo (LMC) builds on RMHMC by using a random velocity in place of RMHMC's random momentum. LMC's explicit integrator is no longer volume preserving; it therefore requires Jacobian corrections for each accept-reject step \cite{lan2015markov}. The embedding geodesic Monte Carlo (egMC) \cite{byrne2013geodesic} is able to take the geometry of the parameter space into account while avoiding implicit integration by splitting the Hamiltonian \cite{shahbaba2014split} into a Euclidean and a geodesic component.  Unfortunately, egMC is not applicable when a manifold's Riemannian embedding is unknown. gLMC, on the other hand, efficiently uses the same split Hamiltonian formulation as egMC but does not require an explicit Riemannian embedding \cite[See][for example]{lan2014spherical}. This last fact makes gLMC an ideal candidate for Bayesian inference on the space of PD matrices. We refer to the algorithm as positive definite Hamiltonian Monte Carlo (PDHMC).

 PDHMC allows us to treat the entire covariance matrix as one would treat any other model parameters.  We are no longer restricted to use a conjugate prior distribution or to specify a low-rank structure. We illustrate applications of PDHMC using both simulated and real-world data. First, we show that PDHMC provides the same empirical results as the closed-form solution for the conjugate Gaussian inverse-Wishart model. After this, we focus on applying Hermitian PDHMC to multivariate spectral density estimation and compare the results obtained from two different prior specifications: the inverse-Wishart prior and the complex reference prior.  Then, we get credible intervals for the squared coherences (see Section \ref{motiv}) of simulated vector auto-regressive time series for which the spectral density matrix is known. Finally, we apply PDHMC to learn the spectral density matrix associated with multivariate local field potentials (LFPs).

\section{Motivation: learning the spectral density matrix}\label{motiv}

Given a stationary multivariate time series $y (t)=(y_1(t), \dots, y_d(t))^{\mathsf T} \in \mathbb{R}^d,\,  t= 1, \dots, T$, one often wants to characterize the dependencies between vector elements through time. There are multiple ways to define such dependencies, and these definitions feature either the time series directly or the Fourier-transformed series in the frequency domain. In the time domain, one characterization of dependence is provided by the lagged variance-covariance matrix $\Gamma_\ell$. In terms of lag $\ell$, this is written
\begin{equation}
\Gamma_\ell = \mbox{Cov}\big(y(t),\, y(t-\ell)\big) = \mbox{E}\Big( \big(y(t)-\mu\big) \big(y(t-\ell) - \mu\big)^T \Big) \ .
\end{equation}
Note that $\Gamma_\ell$ and $\mu$ are invariant over time by stationarity. If the scientist has a reason to suspect that a certain lag $\ell$ is important \emph{a priori}, then $\Gamma_\ell$ can be a useful measure.  On the other hand, it is often more scientifically tractable to think in terms of frequencies rather than lags. In neuroscience, for example, one might hypothesize that two brain regions have `correlated' activity during the performance of a specific task, but this co-activity may be too complex to describe in terms of a simple lagged relationship. The spectral density approach lends itself naturally to this kind of question. For a full discussion, see \cite{ombao2016}.

The power spectral density matrix is the Fourier transform of $\Gamma_\ell$:
\begin{equation}
\Sigma (\omega ) = \sum_{\ell=-\infty}^{\infty} \Gamma_\ell \, \exp (- i 2 \pi \omega \ell) \ .
\end{equation}
$\Sigma(\omega)$ is a Hermitian PD matrix.  A Hermitian matrix $M$ is a complex valued matrix satisfying $M^H=\overline M^{\mathsf T}= M$, where $\overline{(\cdot)}$ denotes taking the complex conjugate. A Hermitian matrix $M$ is defined to be PD if $z^H M z >0,\, \forall z \in \mathbb C^d\setminus \{0\}$. 

A diagonal element $\Sigma_{ii}(\omega)$ is called the auto-spectrum of $y_i(t)$ at frequency $\omega$, and an off-diagonal element $\Sigma_{ij}(\omega),\, i\neq j$ is the cross-spectrum of $y_i(t)$ and $y_j(t)$ at frequency $\omega$. The squared coherence is given by
\begin{equation}
\rho^2_{ij}(\omega) = \frac{|\Sigma_{ij}(\omega)|^2}{\Sigma_{ii}(\omega) \, \Sigma_{jj}(\omega)} \ ,
\end{equation}
where $|\cdot |$ denotes the complex modulus. There are a number of ways to estimate the spectral density matrix and, hence, the matrix of squared coherences. In this paper, we model the discrete Fourier transformed time series $Y(\omega_k) \in \mathbb{C}^d$ as following a complex multivariate Gaussian distribution:
\begin{equation}
Y(\omega_k) \overset{ind}{\sim} \mbox{CN}_d(0, \Sigma(\omega_k)) \ ,
\end{equation}
where, for $\omega_k = \frac{k}{T}$ and $k = - (\frac{T}{2}-1 ),\dots, \frac{T}{2}$, 
\begin{equation}
Y(\omega_k ) = \frac{1}{\sqrt{T}}\sum_{t=1}^{T} y(t) \, \exp (- i 2 \pi \omega_k t ) \ .
\end{equation}
Three assumptions are made here. First, we assume that the $Y(\omega_k)$s are \emph{exactly} Gaussian: this is true when the $y(t)$ follow any Gaussian process. Moreover, if $y(t)$ follow a linear process, then the $Y(\omega_k)$ are asymptotically Gaussian as $T$ goes to infinity \cite{brockwell2013time}. Second, we assume that for $\omega_k \neq \omega_{k'}$, $Y(\omega_k)$ and $Y(\omega_{k'})$ are independent, whereas \cite{brockwell2013time} show that they are asymptotically uncorrelated. 
Third, we assume that $\Sigma (\cdot)$ varies smoothly and slowly as a function over the frequency domain, and take all $Y(\omega_k)$ to be approximately i.i.d. within a small enough frequency band. For example, if we are interested in the alpha band---the interval of frequencies ranging from 7.5 to 12.5 Hz---then we model
\begin{equation}
Y(\omega_k) \overset{iid}{\sim} \mbox{CN}_d(0, \Sigma_\alpha) \ ,
\end{equation}
where $\Sigma_\alpha$ denotes the spectral density matrix shared by the entire band.

Thus, having obtained samples $Y(\omega)$ from a fixed frequency band, we will use Hermitian PDHMC to perform inference on $\Sigma$. The posterior samples of $\Sigma$ automatically provide samples for the distributions of the squared coherences, which can in turn elucidate dependencies between the univariate time series. Before discussing PDHMC, we establish necessary facts regarding the space of positive definite matrices.

\section[The space of positive definite matrices]{The space of positive definite matrices\footnotemark[2] }\label{PDspace}

\footnotetext[2]{In this section we focus on the space of Hermitian positive definite matrices, since the class of symmetric matrices belongs to the broader class of Hermitian matrices.  If the reader is primarily interested in the smaller class, then she is free to substitute $\mathbb{R}$ for $\mathbb{C}$, transpose $(\cdot)^{\mathsf T}$ for conjugate transpose $(\cdot)^H$, and the orthogonal group $O(d)$ for the unitary group $U(d)$.}

Let $\mathcal{S}_d(\mathbb{C})$ denote the space of $d\times d$ Hermitian matrices, and $\mathcal{S}^+_d(\mathbb{C})$ denote its subspace of PD matrices.
The space of Hermitian PD matrices, $\mathcal{S}^+_d(\mathbb{C})$, may be written as a quotient space $GL(d,\mathbb{C})/U(d)$ of the complex general linear group $GL(d,\mathbb{C})$ and the unitary group $U(d)$. The general linear group is the smooth manifold for which every point is a matrix with non-zero determinant.  The unitary group is the space of all complex matrices $U$ satisfying $U^HU=UU^H=I$. This quotient space representation is rooted in the fact that every PD matrix may be written as the product $\Sigma = GG^H = GUU^HG^H$ for a unique $G \in GL(d,\mathbb{C})$ and any arbitrary unitary matrix $U \in U(d)$. For the convenience of exposition, we drop the dependence on $\mathbb C$ of symbols in the following of this section. Related references are \cite{pennec2006riemannian, moakher2011riemannian, helgason1979differential}.

$\mathcal{S}^+_d$ is a homogeneous space with respect to the general linear group: this means that the group acts transitively on the $\mathcal{S}^+_d$. Here the group action is given by conjugation: 
\begin{equation}\label{eq::grpact}
G^* \Sigma = G \Sigma G^H \ .
\end{equation}
For any $\Sigma_1, \Sigma_2 \in \mathcal{S}^+_d$, it simply takes the composition $\Sigma_2^{1/2*} \circ \Sigma_1^{-1/2*}$ to move from $\Sigma_1$ to $\Sigma_2$. 
The space of Hermitian matrices, $\mathcal{S}_d$, happens to be the tangent space to the space of Hermitian PD matrices at the identity, denoted as $T_{Id}\mathcal{S}^+_d$, that is,
$T_{Id}\mathcal{S}^+_d = \mathcal{S}_d$.
The action 
\begin{equation}
\Sigma^{1/2*}: V \mapsto \Sigma^{1/2}V\Sigma^{1/2}
\end{equation}
translates vector $V \in T_{Id}\mathcal{S}^+_d$ to its corresponding vector in $T_{\Sigma}\mathcal{S}^+_d$, the tangent space to the space of PD matrices at point $\Sigma$.

\'Elie Cartan constructed a natural Riemannian metric $g(\cdot,\cdot)$ on the tangent bundle $T \mathcal{S}^+_d$ that is invariant under group action \eqref{eq::grpact}. For two vectors $V_1, V_2 \in T_{Id}\mathcal{S}^+_d$, the metric is given by
\begin{equation}
g_I(V_1, V_2) = \mbox{tr} (V_1 V_2) \ .
\end{equation}
In this way the space of PD matrices is isometric to Euclidean space (equipped with the Frobenius norm) at the identity. Next define the metric at any arbitrary point $\Sigma$ to be 
\begin{equation}
g_\Sigma(V_1,V_2) = \mbox{tr} (\Sigma^{-1} V_1 \Sigma^{-1} V_2)
\end{equation}
It is easy to check that $g_{I}(V_1,V_2) = g_{\Sigma} (\Sigma^{1/2*}V_1, \Sigma^{1/2*}V_2)$ and so $\Sigma^{1/2*}$ is a Riemannian isometry on $\mathcal{S}^+_d$.

Two geometric quantities are required for our purposes: the Riemannian metric tensor and its corresponding geodesic flow, specified by a starting point and an initial velocity vector. The computational details involving the metric tensor are presented in Section \ref{pdhmc}. Here we present the closed form solution for the geodesic flow as found in \cite{pennec2006riemannian}. $\mathcal{S}^+_d$ is an affine symmetric space \cite{koh1965affine}.  As such the geodesics under the invariant metric are generated by the one-parameter subgroups of the acting Lie group \cite{pennec2006riemannian,helgason1979differential}. These one-parameter subgroups are given by the group exponential map which, at the identity, is given by the matrix exponential $\exp tG$.  In order to calculate the unique geodesic curve with starting position $\Sigma (0)$ and initial velocity $V (0)$, all one needs is to translate the velocity to the identity, compute the matrix exponential, and translate it back to the point of interest.  In sum, the geodesic is given by
\begin{eqnarray} \label{eq::geo1}
\Sigma (t) = \exp_\Sigma tV(0) = \Sigma(0)^{1/2} \exp \Big( t \Sigma(0)^{-1/2} V(0) \Sigma(0)^{-1/2} \Big) \Sigma(0)^{1/2} \, .  
\end{eqnarray}
The corresponding flow on the tangent bundle will also be useful. This is obtained by taking the derivative with respect to $t$:
\begin{eqnarray} \label{eq::geo2}
V(t) = \dot\Sigma (t) = \frac{d}{d t} \exp_\Sigma tV(0) = V(0) \Sigma(0)^{-1/2} \exp \Big( t \Sigma(0)^{-1/2} V(0) \Sigma(0)^{-1/2} \Big) \Sigma(0)^{1/2} \, .  
\end{eqnarray}
For any lie group, the exponential map (on which the above formula is based) is a local diffeomorphism between the tangent space at a point on the manifold and the manifold itself.  Given a tangent vector $V$ at $\Sigma$, $\exp_\Sigma V$ is a point on the manifold. Incidentally, for the spaces of PD matrices, this diffeomorphism is global.  The inverse of the exponential map is the logarithmic map.  Whereas the exponential map on the manifold takes Hermitian matrices ($\mathcal{S}_d$) to Hermitian PD matrices ($\mathcal{S}^+_d$), the logarithmic map takes Hermitian PD matrices ($\mathcal{S}^+_d$) to Hermitian matrices ($\mathcal{S}_d$).

Together these are most of the geometric quantities required for PDHMC.  The next section presents Hamiltonian Monte Carlo, its geometric extension RMHMC, and its Lagrangian manifestations.

\section{Bayesian inference using the geodesic Lagrangian Monte Carlo}

Given data $y_1, \dots, y_N \in \mathbb{R}^n$, one may specify a generative model by a likelihood function, $p(y | q)$. 
In the following we allow $q \in \mathcal{M}^m$ to be an m-dimensional vector on a manifold that parameterizes the likelihood. Endowing $q$ with a prior distribution $p(q)$ renders the posterior distribution
\begin{equation}
\pi(q)=p(q | y) = \frac{p(y|q)p(q)}{\int p(y|q)p(q)\, \mbox{d}q} \; .
\end{equation}
The integral is often referred to as the evidence and may be interpreted as the probability of observing data $y$ given the model. In most interesting models the evidence integral is intractable and high dimensional models do not lend themselves to easy numerical integration. Non-quadrature sampling techniques such as importance sampling or even random walk MCMC also suffer in high dimensions.  HMC is an effective sampling tool for higher dimensional models over continuous parameter spaces \cite{duane1987hybrid,neal2011mcmc}. Here we discuss HMC and its geometric variants (see Section \ref{intro}) in detail.

In HMC, a Hamiltonian system is constructed that consists of the parameter vector $q$ and an auxiliary vector $p$ of the same dimension.  The negative-log transform turns the probability density functions into a potential energy function $U(q)=-\log \pi(q)$ and corresponding kinetic function $K(p)$.  Thus $q$ and $p$ become the position and momentum of Hamiltonian function
\begin{equation}
H(q,p) = U(q) + K(p) \; .
\end{equation}
By Euler's method or extensions, the system is numerically advanced according to Hamilton's equations:
\begin{eqnarray}
\frac{dq}{dt} &=& \frac{\partial H}{\partial p} \\ \nonumber
\frac{dp}{dt} &=& -\frac{\partial H}{\partial q}  \; .
\end{eqnarray}  
Riemannian manifold HMC uses a slightly more complicated Hamiltonian to sample from posterior $\pi(q)$:
\begin{eqnarray}
H(q,p) &=& - \log \pi(q) + \frac{1}{2} \log |G(q)| + \frac{1}{2} p^{\mathsf T} G(q)^{-1} p \;.
\end{eqnarray}
Here, $G(q)$ is the Fisher information matrix at point $q$ (in Euclidean space) and may be interpreted as a Riemannian metric tensor induced by the curvature of the log-probability. Exponentiating and negating $H(q,p)$ reveals $p$ to follow a Gaussian distribution centered at origin with metric tensor $G(q)$ for covariance.  The corresponding system of first-order differential equations is given by
\begin{eqnarray}
\frac{dq}{dt} &=&  G(q)^{-1} p, \\ \nonumber
\frac{dp}{dt} &=&  \nabla_q \Big( \log \pi (q) - \frac{1}{2} \log |G(q)|  - \frac{1}{2} p^{\mathsf T} G(q)^{-1} p \Big) \; .
\end{eqnarray}  
The Hamiltonian is not separable in $p$ and $q$. To get numerical solutions, one may split it into a potential term $H^{[1]}$, featuring $q$ alone, and a kinetic term, $H^{[2]}$, featuring both variables \cite{shahbaba2014split, byrne2013geodesic}.  The two systems are then simulated in turn.  The first term is given by
\begin{equation}\label{eq::H1}
H^{[1]}(q, p) = - \log \pi(q) + \frac{1}{2} \log |G(q)|
\end{equation}
 and starting at $(q(0),p(0))$ the associated system has solutions
\begin{eqnarray} \label{eq::update1}
q(t)=q(0) & \mbox{and} & p(t) = p(0) + t \nabla_q\big( \log \pi (q)- \frac{1}{2} \log |G(q)|\big)  |_{q=q(0)} \; .
\end{eqnarray}
The second  component is the quadratic form 
\begin{equation} \label{eq::H2}
H^{[2]} (q,p) = \frac{1}{2} p^{\mathsf T} G(q)^{-1} p
\end{equation}
The solutions to the system associated with H$^{[2]}$ are given by the geodesic flow under the Levi-Civita connection with respect to metric $G$ and with momentum $p(t) = G(q(t))\dot{q}(t)$.  There is, however, no \emph{a priori} reason to restrict $G(q)$ to be the Fisher information as is done in the \cite{girolami2011riemann}. In fact, by allowing $G(q)$ to take on other forms, one may perform HMC on a number of manifold parameterized models.

\subsection{Geodesic Lagrangian Monte Carlo} \label{gLMC}

\begin{algorithm}[t] 
  \caption{Geodesic Lagrangian Monte Carlo} \label{alg::gLMC} 
  \begin{algorithmic}[1]
   \State $v \sim N(0,G^{-1}(q))$
   \State $e \gets - \log \pi (q) - \frac{1}{2} \log |G(q)| +  \frac{1}{2}v^{\mathsf T} G(q) v$
   \State $q^* \gets q$
   \For{$\tau = 1,\dots, T$}
   	\State $v \gets v + \frac{\epsilon}{2} G(q^*)^{-1} \nabla_q \big( \log \pi (q^*) + \frac{1}{2} \log |G(q^*)|\big)$
	\State \parbox[t]{\dimexpr\linewidth-3em}{Progress $(q^*,v)$ along the geodesic flow for time $\epsilon$.\strut}
   	\State $v \gets v + \frac{\epsilon}{2} G(q^*)^{-1} \nabla_q \big( \log \pi (q^*) + \frac{1}{2} \log |G(q^*)|\big)$
   \EndFor
   \State $e^* \gets - \log \pi (q^*) - \frac{1}{2} \log |G(q^*)| +  \frac{1}{2}v^{\mathsf T} G(q^*) v$
    \State $u \sim U(0, 1)$
    \If{$u < \exp (e-e^*)$ }
    	\State $q \gets q^*$
    \EndIf
  \end{algorithmic}  
\end{algorithm}

\cite{byrne2013geodesic} show how to extend the RMHMC framework to manifolds that admit a known Riemannian isometric embedding into Euclidean space. The algorithm is especially efficient when there exists a closed form linear projection of vectors in the ambient space onto the tangent space at any point.  Although this embedding will always exist \cite{nash1956imbedding}, it is rarely known. When equipped with the canonical metric, the space of PD matrices does not admit a known isometric embedding.  Moreover, we are unaware of a closed-form projection onto the manifold's tangent space at a given point. We therefore opt for an intrinsic approach instead.
 
In the prior section, we stated that the solution to Hamilton's equations associated with the kinetic term $H^{[2]}$ is given by the geodesic flow with respect to the Levi-Civita connection. This flow is easily written in terms of the exponential map with respect to a velocity vector (as opposed to the momentum covector). Given an arbitrary covector $p \in T_q^*\mathcal{M}$, one may obtain the corresponding vector $v \in T_q\mathcal{M}$ by the one-to-one transformation $v=G^{-1}(q)p$. Hence whereas RMHMC augments the system with $p \sim N(0, G(q))$, Lagrangian Monte Carlo makes use of $v= G^{-1}(q)p \sim N(0,G^{-1}(q))$.  The energy function is then given by
\begin{eqnarray}\label{eq::energy}
E(q,v) &\propto& - \log \pi (q) - \frac{1}{2} \log |G(q)| +  \frac{1}{2}v^{\mathsf T} G(q) v \:.
\end{eqnarray}
The probabilistic interpretation of the energy remains the same as in the case of RMHMC: the energy is the negative logarithm of the probability density functions of two independent random variables, one of which is the variable of interest, the other of which is the augmenting Gaussian variable. On the other hand, the physical interpretation is different. We use the term `energy' in order to accommodate the two physical interpretations available for \eqref{eq::energy}: $E(q,v)$ may be thought of either as a Hamiltonian or as a Lagrangian energy. In practice, which formulation is used is dictated by the geometric information available. 
The Lagrangian formulation provides efficient update equations when no closed-form geodesics are available. In this case, the Lagrangian (energy) is defined as the kinetic term $T$ less the potential term $V$ as follows
\begin{eqnarray}
V(q,v) = \log \pi (q) + \frac{1}{2} \log |G(q)|, &\mbox{and}& T(q,v) = \frac{1}{2}v^{\mathsf T} G(q) v  \; .
\end{eqnarray}
But when closed-form geodesics are available, it is useful to follow \cite{byrne2013geodesic} and split the (now considered) Hamiltonian into two terms as in \eqref{eq::H1} and \eqref{eq::H2}. Within this regime, $H^{[1]}= - V$ and $H^{[2]}=T$.
In analogy with \eqref{eq::update1} and starting at $(q(0),v(0))$, the system defined by potential $V$ has solution
\begin{eqnarray}\label{eq::gLMCupdate}
q(t)=q(0), && v(t) = v(0) + t \, G(q)^{-1} \nabla_q V (q,v) |_{q=q(0)} \;,
\end{eqnarray}
and the system defined by kinetic term $T$ has the unique geodesic path specified by starting position $q(0)$ and initial velocity $v(0)$ as a solution. The inverse metric tensor $G^{-1}(q)$ is used to `raise the index', i.e. transform the covector $\nabla_q V (q,v)$ into a vector on the tangent space at $q$.  Thus it plays a similar function to the orthogonal projection in \cite{byrne2013geodesic}. We call this formulation geodesic Lagrangian Monte Carlo (gLMC) and detail its steps in Algorithm \ref{alg::gLMC}, where the term `Lagrangian' is used to emphasize the fact that we use velocities in place of momenta.
Note \cite{lan2014spherical,Lan2016} implemented the similar idea on the manifold of a $d$-dimensional sphere.
To implement geodesic Lagrangian Monte Carlo, one must be able to compute the inverse metric tensor $G^{-1}(q)$ and the geodesic path given starting values. When the space of PD matrices is equipped with the canonical metric, $G^{-1}(q)$ is given in closed-form and the geodesic path is easily computable.

\section{Positive definite Hamiltonian Monte Carlo}\label{pdhmc}

\begin{algorithm}[t]
  \caption{Symmetric PDHMC}  
  \begin{algorithmic}[1] 
   \State $\mbox{vech}(V) \sim N(0, G^{-1}(\Sigma_t))$
   \State $e \gets - \log \pi (\Sigma_t) - \frac{d+1}{2} \log |\Sigma_t|  +  \frac{1}{2} \mbox{vech}(V) ^{\mathsf T} G(\Sigma_t) \mbox{vech}(V)$
   \State $\Sigma^* \gets \Sigma_t$
   \For{$\tau = 1,\dots, T$}
   	\State $ \mbox{vech}(V) \gets \mbox{vech}(V) + \frac{\epsilon}{2} G^{-1}(\Sigma^*)\mbox{vech}\Big( \nabla_{\Sigma}\big( \log \pi (\Sigma^*)+  \frac{d+1}{2} \log |\Sigma^*| \big) \Big)$
	\State Progress $(\Sigma^*,V)$ along the geodesic flow for time $\epsilon$.
   	\State $ \mbox{vech}(V) \gets \mbox{vech}(V) + \frac{\epsilon}{2} G^{-1}(\Sigma^*)\mbox{vech}\Big( \nabla_{\Sigma}\big( \log \pi (\Sigma^*)+  \frac{d+1}{2} \log |\Sigma^*| \big) \Big)$
   \EndFor
\State $e^* \gets - \log \pi (\Sigma^*) - \frac{d+1}{2} \log |\Sigma^*|  +\frac{1}{2} \mbox{vech}(V) ^{\mathsf T} G(\Sigma^*) \mbox{vech}(V)$
    \State $u \sim U(0, 1)$
    \If{$u < \exp (e -e^*)$ }
    	\State $\Sigma_{t+1} \gets \Sigma^*$
    \EndIf
  \end{algorithmic}\label{alg1}
\end{algorithm}

Positive definite Hamiltonian Monte Carlo (PDHMC) is geodesic Lagrangian Monte Carlo on the space of PD matrices, $\mathcal{S}^+_d$, equipped with the canonical metric. In order to signify that we are no longer dealing with gLMC in its full generality, we adopt the notation of Section \ref{PDspace}. PD matrix $\Sigma$ replaces $q$, and symmetric or Hermitian matrix $V$ replaces $v$. All other notations remain the same.  As stated in the previous section, we require the inverse metric tensor $G^{-1}(\Sigma)$. To compute this quantity, we need a couple more tools provided by \cite{moakher2011riemannian}.  Let $\mbox{vech}(\cdot)$ take symmetric (Hermitian) $d\times d$ matrices to vectors of length $\frac{d}{2}(d+1)$ by stacking diagonal and subdiagonal matrix elements in the following way:
\begin{eqnarray}
\mbox{vech}(V) = (V_{11}, V_{21},\dots, V_{d1}, V_{22}, \dots, V_{d2}, \dots, V_{dd}).
\end{eqnarray}
Let $\mbox{vec}(\cdot)$ take symmetric (Hermitian) $d\times d$ matrices to vectors of length $d^2$ by stacking all matrix elements:
\begin{eqnarray}
\mbox{vec}(V) = (V_{11}, V_{21},\dots, V_{d1}, V_{12}, \dots, V_{d2}, \dots,V_{1d},\dots, V_{dd}).
\end{eqnarray}
Let $D_d$ be the unique $d^2 \times \frac{d}{2}(d+1)$ matrix satisfying
\begin{eqnarray}
\mbox{vec}(V) = D_d \mbox{vech}(V)\; .
\end{eqnarray}
Denote $D_d^+$ as the Moore-Penrose inverse of $D_d$ satisfying
\begin{eqnarray}
\mbox{vech}(V) = D^+_d \mbox{vec}(V)\; ,
\end{eqnarray}
with $D^+_d$ given by
\begin{eqnarray}
D^+_d = (D_d^{\mathsf T} D_d)^{-1} D_d^{\mathsf T} \; .
\end{eqnarray}
Then \cite{moakher2011riemannian} show that the metric tensor and inverse metric tensor are given by the $\frac{d}{2}(d+1) \times \frac{d}{2}(d+1)$ dimensional matrices
\begin{eqnarray}\label{eq::tensors}
G(\Sigma) = D_d^{\mathsf T}(\Sigma^{-1} \otimes \Sigma^{-1})D_d &\mbox{and}& G^{-1}(\Sigma) = D_d^+(\Sigma \otimes \Sigma)D_d^{+T}\;.
\end{eqnarray}
Finally, the determinant of $G(\Sigma)$ can be expressed in terms of $\Sigma$ alone:
\begin{equation}
|G(\Sigma)| \propto |\Sigma|^{d+1} \ .
\end{equation}

The metric tensor features in the energy function for both symmetric and Hermitian PDHMC.  For symmetric PDHMC, the energy is given by
\begin{align}
E(\Sigma, V) &\propto - \log \pi (\Sigma) - \frac{1}{2} \log |G(\Sigma)|  +\frac{1}{2} \mbox{vech}(V) ^{\mathsf T} G(\Sigma) \mbox{vech}(V) \\ \nonumber
 &\propto - \log \pi (\Sigma) - \frac{d+1}{2} \log |\Sigma|  +\frac{1}{2} \mbox{vech}(V) ^{\mathsf T} G(\Sigma) \mbox{vech}(V)\ ,
\end{align}
but the energy associated with Hermitian PDHMC is slightly different. In this case, both $\Sigma$ and $V$ are complex valued, and vech$(V)$ follows a multivariate complex Gaussian distribution with covariance $G^{-1}(\Sigma)$. Therefore, the Hermitian PDHMC energy is given by
\begin{align}
E(\Sigma, V) &\propto - \log \pi (\Sigma) -  \log |G(\Sigma)|  + \mbox{vech}(V) ^H G(\Sigma) \mbox{vech}(V) \\ \nonumber
&\propto - \log \pi (\Sigma) - (d+1) \log |\Sigma|  + \mbox{vech}(V) ^H G(\Sigma) \mbox{vech}(V)
\end{align}
where $(\cdot)^H$ signifies the conjugate transpose. Notice that the log-determinant and quadradic terms are not multiplied by the factor $1/2$.  This accords with the density function of a complex Gaussian random variable.  See Appendix \ref{A} for more details.

The metric tensor \eqref{eq::tensors} and the geodesic equations \eqref{eq::geo1} and \eqref{eq::geo2} are the only geometric quantities required for PDHMC. The $t$-th iteration of the symmetric PDHMC algorithm is shown in Algorithm \ref{alg1}. The $t$-th iteration of the Hermitian PDHMC algorithm is shown in Algorithm \ref{alg2}. First, one generates a Gaussian initial velocity on $T_{\Sigma_t} \mathcal{S}^+_d$ (Step 1). Then, the energy function is evaluated and stored (Step 2). Next, the system is numerically advanced using the split Hamiltonian scheme.  Following Equation \eqref{eq::gLMCupdate}, the velocity vector $V$ is updated one half-step with the gradient of H$^{[1]}$ (Step 4). For Step 5, both $\Sigma$ and $V$ are updated with respect to $H^{[2]}$, i.e. they are transported along the geodesic flow given by Equations \eqref{eq::geo1} and \eqref{eq::geo2}:
\begin{equation}
\big[ \Sigma(0), V(0) \big] \mapsto \big[ \Sigma(\epsilon), V(\epsilon) \big] \ .
\end{equation}
Again, the velocity vector $V$ is updated one half-step with the gradient of H$^{[1]}$ (Step 6). Finally, the energy is evaluated at the new Markov state (Step 9), and a Metropolis accept-reject step is implemented (Steps 10-12).  It is important to note that, besides being over different algebraic fields, the symmetric and Hermitian PDHMC only differ in their respective energies. The general implementation is the same.  See Appendix \ref{A} for a short discussion on gradients.

\begin{algorithm}[t]
  \caption{Hermitian PDHMC}  
  \begin{algorithmic}[1]
   \State $\mbox{vech}(V) \sim \mbox{CN}(0, G^{-1}(\Sigma_t))$
   \State $e \gets - \log \pi (\Sigma_t) - (d+1)\log |\Sigma_t|  + \mbox{vech}(V) ^H G(\Sigma_t) \mbox{vech}(V)$
   \State $\Sigma^* \gets \Sigma_t$
   \For{$\tau = 1,\dots, T$}
   	\State $ \mbox{vech}(V) \gets \mbox{vech}(V) + \frac{\epsilon}{2} G^{-1}(\Sigma^*)\mbox{vech}\Big( \nabla_{\Sigma}\big( \log \pi (\Sigma^*)+  (d+1) \log |\Sigma^*| \big) \Big)$
	\State Progress $(\Sigma^*,V)$ along the geodesic flow for time $\epsilon$.
   	\State $ \mbox{vech}(V) \gets \mbox{vech}(V) + \frac{\epsilon}{2} G^{-1}(\Sigma^*)\mbox{vech}\Big( \nabla_{\Sigma}\big( \log \pi (\Sigma^*)+  (d+1) \log |\Sigma^*| \big) \Big)$
   \EndFor
\State $e^* \gets -\log \pi (\Sigma^*) - (d+1) \log |\Sigma^*|  + \mbox{vech}(V) ^H G(\Sigma^*) \mbox{vech}(V)$
    \State $u \sim U(0, 1)$
    \If{$u < \exp (e -e^*)$ }
    	\State $\Sigma_{t+1} \gets \Sigma^*$
    \EndIf
  \end{algorithmic}\label{alg2}
\end{algorithm}

\section{Results}

This section features empirical validation of the PDHMC algorithm as well as an application to learning the spectral density matrix for vector time series.  For empirical validation, we present quantile-quantile and trace plots comparing the PDHMC sample to the closed-form solution made available by the conjugate prior. We then use Hermitian PDHMC to learn the spectral density matrices of both simulated and LFP time series.  We use the posteriors thus obtained to get credible intervals on the squared coherences for the vector time series.

\subsection{Empirical validation}

 \begin{figure}[t]
    \centering
    \includegraphics[width=\textwidth]{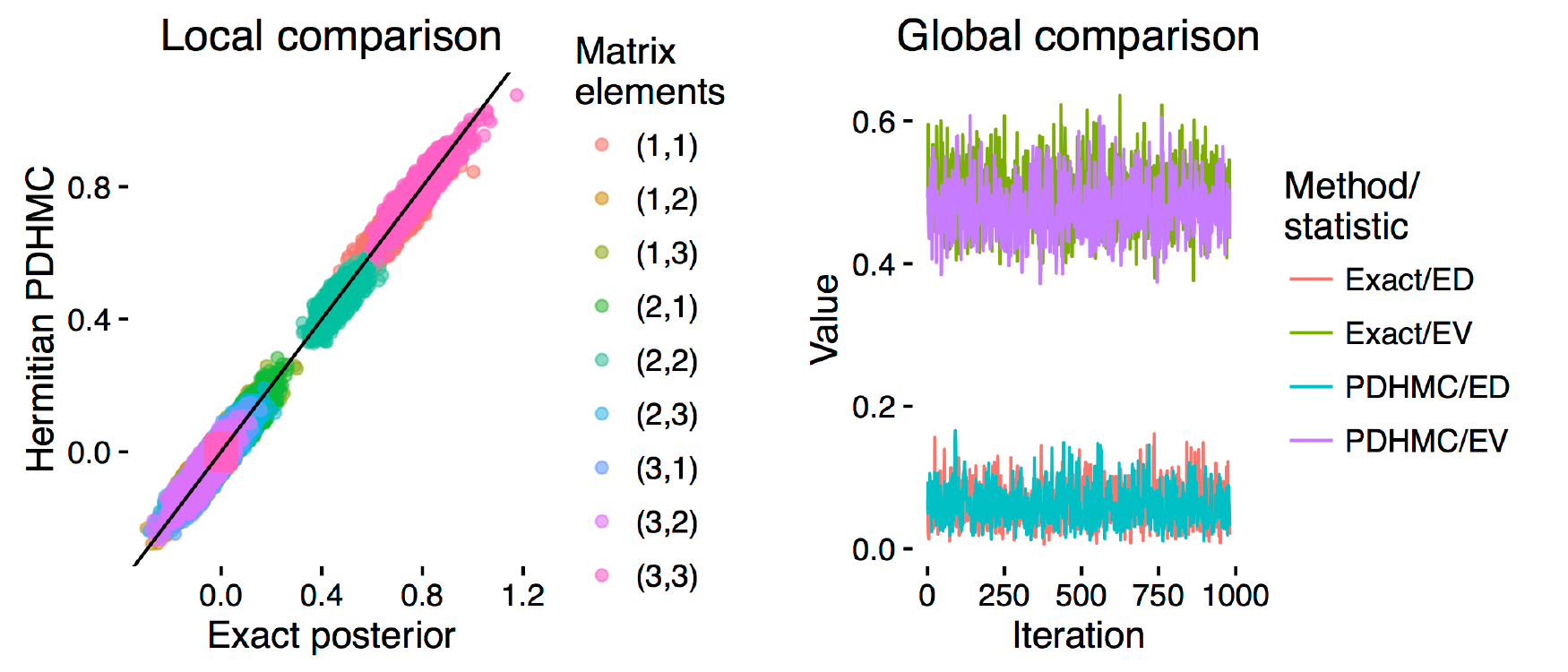}
  \caption{These figures provide empirical validation for the well-posedness of PDHMC. On the left is a quantile-quantile plot comparing the Hermitian PDHMC posterior sample with that of the closed-form posterior for the complex Gaussian inverse-Wishart model. Both real and imaginary elements are included, and points are jittered for visibility. On the right are posterior samples of `global' matrix statistics pertaining both to (symmetric) PDHMC and the closed-form solution. These statistics are the effective variance and the effective dependence, built off the covariance matrix and the correlation matrix, respectively.}
  \label{fig::valid_plot}
 \end{figure}
 
  \begin{figure}[t]
    \centering
    \includegraphics[width=\textwidth]{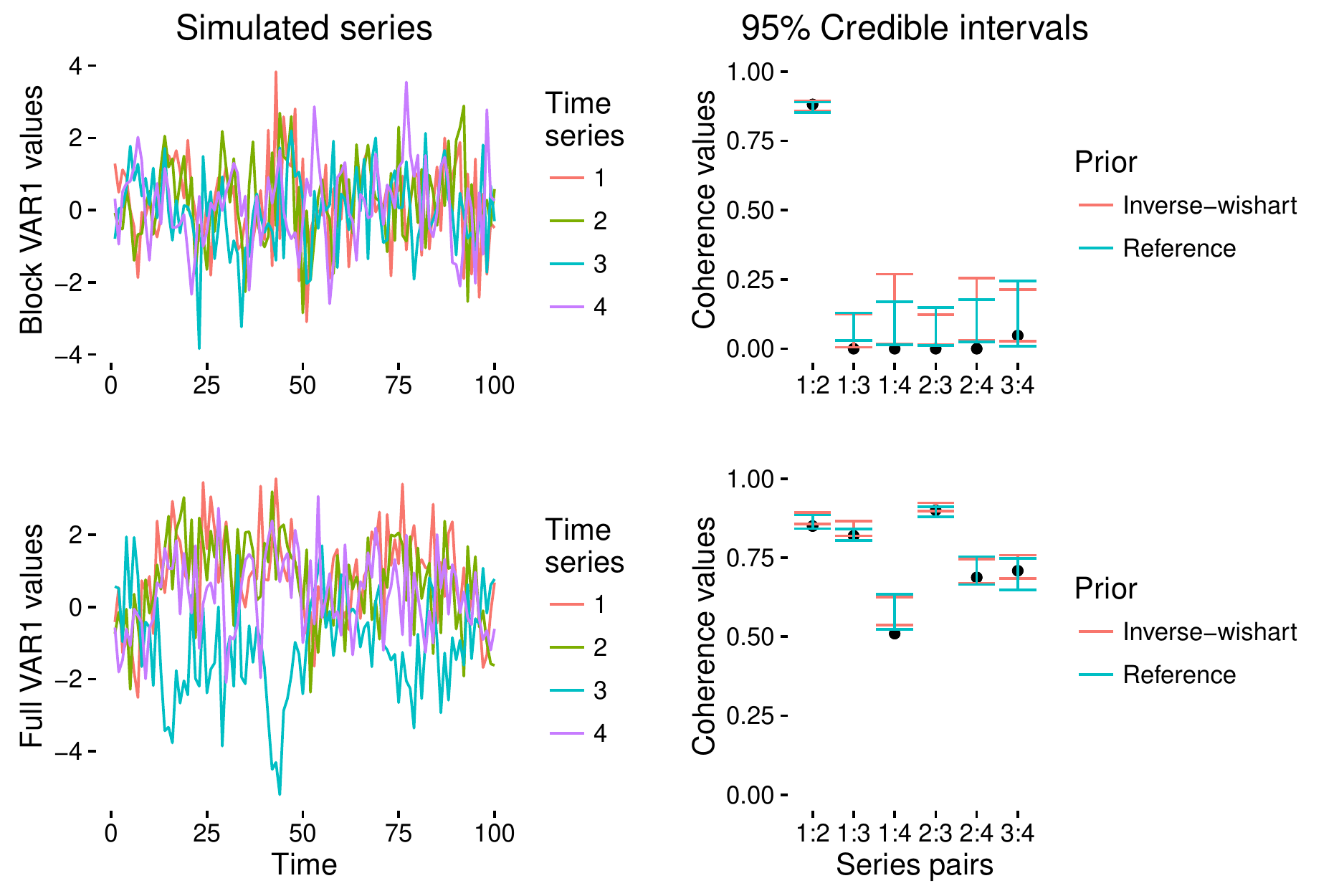}
  \caption{Two 4-dimensional VAR1 time series and credible intervals for their 6 corresponding coherences measured at 20-40 Hz: the top row belongs to a block VAR1 process characterized by two independent 2-dimensional VAR1 time series; the bottom row belongs to a full VAR1 process. The left column shows the first 100 samples of both time series, each of which totals 5,000 samples in length.  The right column shows credible intervals from posteriors obtained using the inverse-Wishart and reference priors.}
  \label{fig::var1_plot}
 \end{figure}

 \begin{figure}[t]
    \centering
    \includegraphics[width=\textwidth]{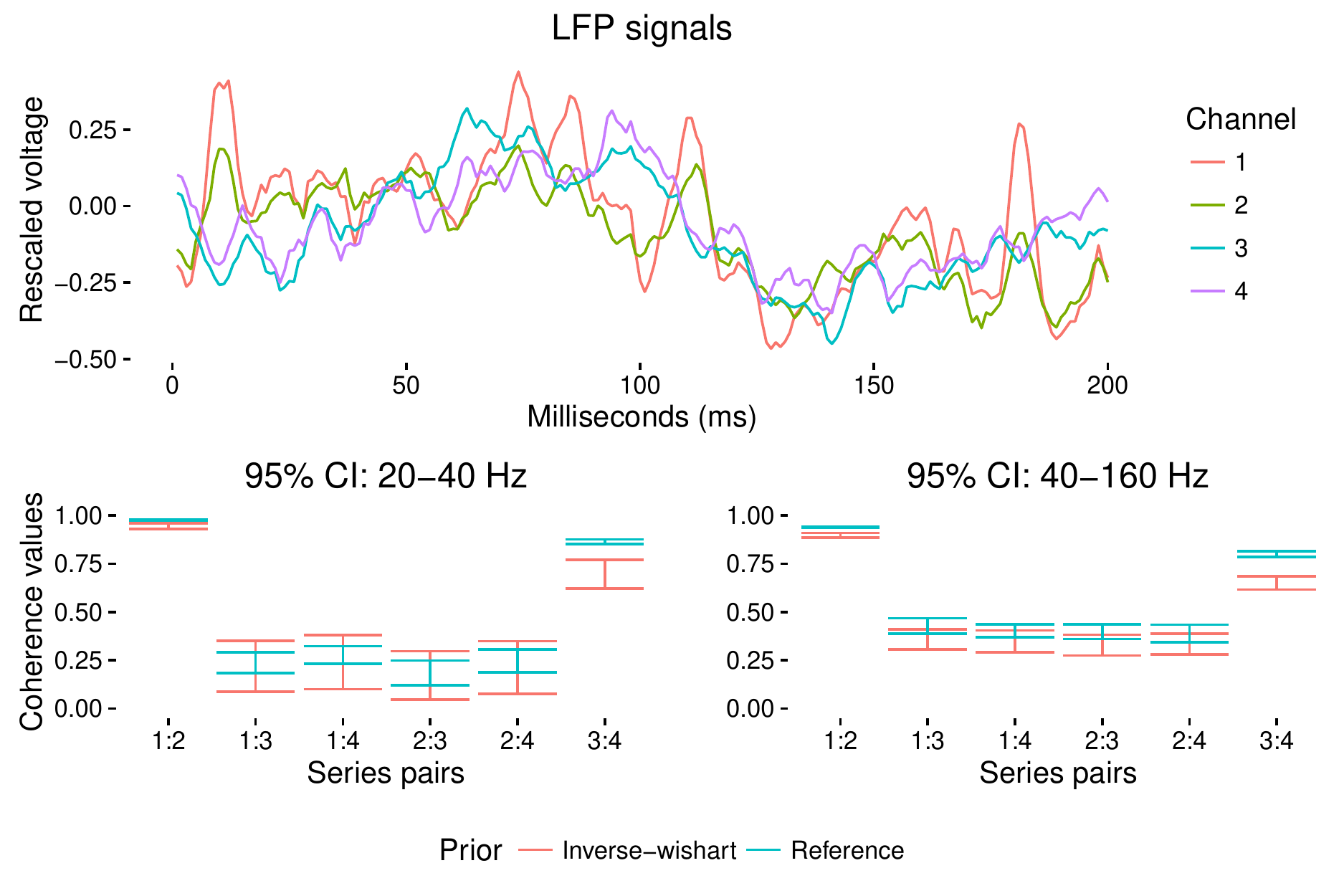}
  \caption{A 4-dimensional LFP signal with credible intervals for 6 coherences measured at 20-40 Hz (left) and 40-160 Hz (right).  First 200 samples are shown for ease of visualization; the multi-dimensional time series totals 4,000 samples in length. Coherence profiles are remarkably similar between the two frequency bands considered.}
  \label{fig::Real_TS_plot}
 \end{figure}

Before applying Hermitian PDHMC to spectral density estimation, we demonstrate validity by comparing samples from empirical posterior distributions of the Gaussian inverse-Wishart model obtained by PDHMC and the closed-form solution. We compare element-wise distributions with quantile-quantile plots and whole-matrix distributions with two global matrix measures. The comparisons based on 10,000 samples by the different sampling methods over $3$-by-$3$ PD matrices are illustrated in Figure \ref{fig::valid_plot}. The first 200 samples are discarded, and, for better visualization, every tenth sample is kept.

On the left panel of Figure \ref{fig::valid_plot}, a quantile-quantile plot is used to compare the Hermitian PDHMC posterior sample to the closed-form posterior. Points are jittered for easy visualization, and each color specifies a different matrix element. Note that some colors appear twice: these double appearances correspond to real and imaginary matrix elements. For example, pink appears at zero as well as the upper right of the plot: this color corresponds to a diagonal matrix element since, on account of the matrix being Hermitian, its imaginary part is fixed at zero.  Most importantly, all matrix elements fit tightly around the line $y=x$, suggesting a perfect match in quantiles between empirical distributions.

On the right panel of Figure \ref{fig::valid_plot}, we present samples obtained from two whole-matrix statistics using symmetric PDHMC and the closed-form posterior.  These measures are the effective variance (EV) and the effective dependence (ED):
\begin{eqnarray}
\mbox{EV} (\Sigma) = |\Sigma|^{1/d}\ , &\mbox{and}  & \mbox{ED} (\Sigma) = 1 - | \mbox{corr}(\Sigma)|^{1/d} \ .
\end{eqnarray}
The EV is the geometric mean of the eigenvalues of the matrix $\Sigma$. It provides a dimension free measure of the total variance encoded in the matrix.  The ED gets its name because the determinant of a correlation matrix is inversely related to the magnitude of the individual correlations that make up the off-diagonals. In addition to seeing that element-wise distributions match, one would also like to know that their joint distributions correspond.  The EV and ED are good measures of global matrix features and here provide empirical evidence for the validity of PDHMC.

\subsection{Learning the spectral density}

An important benefit of PDHMC is that it enables practitioners to specify prior distributions other than the inverse-Wishart on PD matrices based on  needs dictated by the problems at hand. PDHMC improves modeling flexibility. We use the problem of Bayesian spectral density estimation to demonstrate the possibility and advantage of using non-conjugate priors. The spectral density matrix $\Sigma(\omega)$ and its coherence matrix $R(\omega)$ are defined in Section \ref{motiv}. In the context of stationary, multivariate time series, the coherences that make up the off-diagonals of $R(\omega)$ provide a lag-free measure of dependence between univariate time series at a given frequency $\omega$.

We compare posterior inference for these coherences between two models with different priors: the first model uses the complex inverse-Wishart prior; the second uses the complex reference prior \cite{svensson2010reference}. The reference prior is an improper prior that has been proposed as an alternative to Jeffrey's prior for its superior eigenvalue shrinkage (which improves asymptotic efficiency of estimators). We use the reference prior to emphasize the flexibility allowed by PDHMC but not as a modeling suggestion. \cite{svensson2010reference} provide a Gibbs sampling routine based on the eigen-decomposition of the covariance matrix. The reference prior's form is provided in Appendix \ref{comp_ref_pri}. 

We apply Hermitian PDHMC to learning the spectral density matrix for three distinct 4-dimensional time series. The first is a simulated first-order vector-autoregressive (VAR1) time series with block structure consisting of two independent, 2-dimensional VAR1 time series. The second time series is also VAR1 but with dependencies allowed between all four of the scalar time series of which it is composed. The third time series comes from local field potentials (LFP) recorded in the CA1 region of a rat hippocampus \cite{allen2016nonspatial}. 

Figure \ref{fig::var1_plot} shows the first 100 samples from both VAR1 time series along with 95\% posterior intervals for the complex moduli of the coherences. The time series are simulated with the form:
\begin{equation}
y(t) = \Phi\, y(t-1) + \epsilon_t,\quad \quad y(1) = \epsilon_1 , \quad \quad \epsilon_t \sim N_4(0,I), \: t=1,\dots,\, 15000 \, ,
\end{equation}
where the eigenvalues of transition matrix $\Phi$ are bounded with absolute value less than 1 to induce stationarity. The first 10,000 data points are discarded to allow time for mixing.  The first row of Figure \ref{fig::var1_plot} belongs to the block VAR1: $\Phi$ is a  randomly constructed, block-diagonal matrix, so the first two scalar time series are independent from the second two. The second row of Figure \ref{fig::var1_plot}  belongs the the second VAR1, all the scalar time series of which are dependent on all the others. Here $\Phi$ is also a randomly constructed matrix but is not block-diagonal. The intervals corresponding to the inverse-Wishart prior model are given in orange. The intervals corresponding to the reference prior model are given in blue. The true coherences are represented by black points and are obtained using the following closed-form formula for the spectral density of a VAR1 process \cite{ombao2016}:
\begin{equation}
\Sigma (\omega) = \big(I-\Phi e^{-2\pi i \omega} \big)^{-1} Q\, \big(I-\Phi e^{-2\pi i \omega} \big)^{-H}.
\end{equation}
Here $Q$ is the covariance matrix of the additive noise $\epsilon_t$, and $(\cdot)^{-H}$ denotes the inverse conjugate transpose. For the block VAR1 example, both models capture the true, non-null coherences (i.e. those given on the far left and the far right), but neither captures the null coherences. This is more than satisfactory, since coherences equal to zero imply the identity for a covariance matrix. By looking closely, one can see that the first and second time series (orange and green) are indeed strongly dependent on each other, as interval `1:2' suggests. For the full VAR1 example, both models capture five out of six true coherences, but the reference prior model gets closer to the truth than the inverse-Wishart model does.  

We use the same tools to detect coherences between LFP signals simultaneously recorded from the CA1 region of a rat hippocampus prior to a memory experiment \cite{allen2016nonspatial}.  Two of the LFP signals are recorded on one end of the CA1 axis, and the other two LFP signals are recorded at the opposite end.   Figure \ref{fig::Real_TS_plot} shows the first 200 of 4,000 samples (recorded at 1,000 Hz) and 95\% credible intervals for the coherences at two different frequency bands: 20-40 Hz and 40-160 Hz. \emph{The spatial discrepancy is reflected in the posterior distributions of the individual coherences.}  Both bands show similar coherence patterns, where spatial location appears to dictate strength of coherence: the leftmost and rightmost pairs are closer to each other in space, while the center pairs are farther from each other.  This reflects what is apparent in the top of Figure \eqref{fig::Real_TS_plot}, where the first and second time series (orange and green) are dependent, and the third and fourth time series (blue and purple) are dependent.  These correspond to the intervals labeled `1:2' and `3:4', respectively.  The credible intervals are smaller for the 20-40 Hz band because that band has only 1/6 the data of the 40-160 Hz band.  Between prior models, the intervals differ more for the 40-160 Hz band.  This is counter-intuitive since the influence of the prior distribution is often assumed to diminish with the size of the data set.  One question is whether this surprising result is related to the reference prior's being the prior that is `maximally dominated by the data' \cite{berger2009formal}. These differences---differences between posterior distributions for the two prior models---communicate that other prior distributions might provide tangible differences between results in spectral analysis and that it would be useful to understand which prior distributions are appropriate in which contexts. 


\section{Discussion}

We presented PDHMC, a Lagrangian Monte Carlo method for Bayesian inference on the space of PD matrices. We outlined its relationship to other geometric extensions of HMC and showed how to apply PDHMC to both symmetric and Hermitian PD matrices. We demonstrated empirical validity using both element-wise and whole-matrix comparisons against the conjugate inverse-Wishart model. Finally, we applied Hermitian PDHMC to Bayesian spectral density estimation. The algorithm proved effective for detecting true coherences of simulated time series, as well as recovering spatial discrepancies between real-world LFP signals.

We see three branches of inquiry stemming from this work: the first is algorithmic; the second, theoretical; the third, methodological. First, what variations of HMC might help extend PDHMC into higher dimensions? There are multiple such extensions that are completely orthogonal to PDHMC. Examples are windowed HMC, geometric extensions to the NUTs algorithm, shortcut MCMC, and look-ahead HMC \cite{neal2011mcmc,  sohl2014hamiltonian, betancourt2013generalizing}.  Auto-tuning will prove useful: even within the same dimension, different samples will dictate different numbers of leapfrog steps and step-sizes. From the theoretical standpoint, the canonical metric on the space of PD matrices is closely related to the Fisher information metric on covariance matrices: how should one characterize this intersection between information geometry and Riemannian symmetric spaces, and how might this relationship inform Bayes estimator properties or future variations on PDHMC? Finally, much work needs to be done in prior elicitation for Bayesian spectral density estimation. Which priors on Hermitian PD matrices should be used for which problems, what are the costs and benefits, and are there priors over symmetric PD matrices that need to be complexified (cf. \cite{schwartzman2015lognormal, fazayeli2016matrix})? A clear delineation will be useful for practitioners in Bayesian time series research.

\newpage
\appendix 

\section{Real and complex matrix derivatives} \label{A}

The derivative of a univariate, real valued function with respect to a matrix is most cleanly calculated using the matrix differential.  This is true whether $f: M_p(\mathbb{R}) \mapsto \mathbb{R}$ or $f: M_p(\mathbb{C}) \mapsto \mathbb{R}$, i.e. whether $f$ is a function over real $p\times p$ matrices or complex $p\times p$ matrices. As an example, we consider the multivariate Gaussian distribution with mean 0 and covariance $\Sigma$.  First, let $f$ be the probability density function over real valued Gaussian random vectors $y_n\in \mathbb{R}^d, n = 1, \dots, N$. Let $Y$ be the $d\times N$ concatenation of these $N$ i.i.d. random variables. Then the log density is given by
\begin{align} \label{eq::loglik}
\log f(Y^N, \Sigma) &\propto - \frac{N}{2} \log |\Sigma| - \frac{1}{2}\sum_{n=1}^N y_n^{\mathsf T} \Sigma^{-1}y_n  \\ \nonumber
&= - \frac{N}{2} \log |\Sigma| - \frac{1}{2}\mbox{tr}\{ \Sigma^{-1}YY^{\mathsf T} \} \ .
\end{align}
We apply the matrix differential to \eqref{eq::loglik} using two general formulas:
\begin{eqnarray} \label{logdetdif}
d\, \log |\Sigma| = \mbox{tr}\{\Sigma^{-1} \ d\, \Sigma\}, & \mbox{and} & d\, \Sigma^{-1} = - \Sigma^{-1}\ (d\, \Sigma) \ \Sigma^{-1} \ ,
\end{eqnarray}
rendering
\begin{align}
d \log f(Y, \Sigma) &= -\frac{N}{2}\mbox{tr}\{ \Sigma^{-1} \ d\, \Sigma \} + \frac{1}{2}\mbox{tr}\big\{ \Sigma^{-1} (d\,\Sigma) \Sigma^{-1} YY^{\mathsf T}\big\}  \\ \nonumber
&=  -\frac{N}{2}\mbox{tr}\{   (d\, \Sigma) \ \Sigma^{-1} \} + \frac{1}{2}\mbox{tr}\big\{(d\,\Sigma) \ \Sigma^{-1}YY^{\mathsf T}  \Sigma^{-1}\big\} \ .
\end{align}
Finally, we relate the matrix differential to the gradient with the fact that, for an arbitrary function $g$,
\begin{eqnarray} \label{diftograd}
d \, g(\Sigma) = \mbox{tr}\{ (d\, \Sigma) A\} & \iff & \nabla_\Sigma\, g(\Sigma) = A \ .
\end{eqnarray}
This gives the final form of the gradient of the log density function with respect to covariance $\Sigma$:
\begin{equation}
\nabla_\Sigma \log f(Y, \Sigma) = -\frac{N}{2}\Sigma^{-1} + \frac{1}{2} \Sigma^{-1} YY^{\mathsf T} \Sigma^{-1} \ .
\end{equation}
For more on the matrix differential, see \cite{magnus1995matrix}. The complex matrix differential is treated in \cite{hjorungnes2007complex} and has a similar form real valued functions. The log density of the multivariate complex Gaussian with mean 0 is given by
\begin{align} \label{eq::loglik2}
\log f(Y, \Sigma) &\propto - N \log |\Sigma| - \sum_{n=1}^N y_n^H \Sigma^{-1}y_n  \\ \nonumber
&= - N \log |\Sigma| - \mbox{tr}\{ \Sigma^{-1}YY^H \} \ ,
\end{align}
where $(\cdot)^H$ denotes the conjugate transpose.  Note that the log density is scaled by a factor of two compared to the real case. The resulting gradient is
\begin{equation}
\nabla_\Sigma \log f(Y, \Sigma) = -N\Sigma^{-1} +  \Sigma^{-1} YY^H \Sigma^{-1} \ .
\end{equation}

\subsection{The complex reference prior}\label{comp_ref_pri}
Gradients of prior probabilities are calculated in a similar way. We demonstrate for the complex reference prior.  Let $\lambda_i, i=1,\dots,d$ be the decreasing eigenvalues of Hermitian PD matrix $\Sigma$. Then the complex reference prior has the following form:
\begin{equation}\label{eq::refpri}
p(\Sigma) \propto \frac{\mbox{d} \Sigma}{|\Sigma|\, \prod_{k<j}(\lambda_k-\lambda_j)^2} \ .
\end{equation}
To use the above approach for deriving the matrix derivatives, we need to be able to write the differential d$\lambda_i$ in terms of the matrix differential d$\Sigma$. \cite{magnus1995matrix} provides the formula when all eigenvalues are distinct:
\begin{equation}
d\lambda_i = \mbox{tr}\Big(\sum_{j=1}^d V^{-1}_{ij} \Sigma^{j-1} d\Sigma \Big)\ ,
\end{equation}
where V is the Vandermonde matrix:
\begin{equation}
V^{\mathsf T} = \begin{vmatrix}
  1 & \lambda_1 & \lambda_1^2 & \cdots & \lambda_1^{n-2} & \lambda_1^{n-1} \\
  1 & \lambda_2 & \lambda_2^2 & \cdots & \lambda_2^{n-2} & \lambda_2^{n-1} \\
\vdots & \vdots & \vdots & \ddots & \vdots & \vdots \\
  1 & \lambda_n & \lambda_n^2 & \cdots & \lambda_n^{n-2} & \lambda_n^{n-1}
\end{vmatrix} \ .
\end{equation}
We now calculate the gradient of the log of the complex reference prior:
\begin{align}
d \log p(\Sigma) &= - d \log |\Sigma| - 2  \sum_{k <j} d\, \log  (\lambda_k - \lambda_j ) \\ \nonumber
& = - \mbox{tr}(\Sigma^{-1} d \Sigma ) - 2 \sum_{k<j} \frac{d\lambda_k-d\lambda_j}{\lambda_k - \lambda_j} \\ \nonumber
&= - \mbox{tr}(\Sigma^{-1} d \Sigma ) - 2 \sum_{k<j} \mbox{tr}\Big(\sum_{i=1}^d \big(V^{-1}_{ki}-V^{-1}_{ji}\big) \Sigma^{i-1} d\Sigma \Big)/ (\lambda_k - \lambda_j) \ .
\end{align}
Combining this with Equations \eqref{logdetdif} and \eqref{diftograd} renders matrix gradient
\begin{equation}
\nabla_\Sigma \log p(\Sigma) \propto - \Sigma^{-1} - 2 \sum_{k<j} \Big(\sum_{i=1}^d \big(V^{-1}_{ki}-V^{-1}_{ji}\big) \Sigma^{i-1} \Big)/ (\lambda_k - \lambda_j) \ .
\end{equation}

\newpage
\bibliographystyle{unsrt}

\begin{thebibliography}{10}

\bibitem{tokuda2011visualizing}
Tomoki Tokuda, Ben Goodrich, I~Van~Mechelen, Andrew Gelman, and F~Tuerlinckx.
\newblock Visualizing distributions of covariance matrices.
\newblock {\em Dept. Statist., Columbia Univ., New York, NY, USA, Tech. Rep},
  2011.

\bibitem{girolami2011riemann}
Mark Girolami and Ben Calderhead.
\newblock Riemann manifold langevin and hamiltonian monte carlo methods.
\newblock {\em Journal of the Royal Statistical Society: Series B (Statistical
  Methodology)}, 73(2):123--214, 2011.

\bibitem{byrne2013geodesic}
Simon Byrne and Mark Girolami.
\newblock Geodesic monte carlo on embedded manifolds.
\newblock {\em Scandinavian Journal of Statistics}, 40(4):825--845, 2013.

\bibitem{lan2014spherical}
Shiwei Lan, Bo~Zhou, and Babak Shahbaba.
\newblock Spherical hamiltonian monte carlo for constrained target
  distributions.
\newblock In {\em JMLR workshop and conference proceedings}, volume~32, page
  629. NIH Public Access, 2014.

\bibitem{neal2011mcmc}
Radford~M Neal et~al.
\newblock Mcmc using hamiltonian dynamics.
\newblock {\em Handbook of Markov Chain Monte Carlo}, 2:113--162, 2011.

\bibitem{lan2015markov}
Shiwei Lan, Vasileios Stathopoulos, Babak Shahbaba, and Mark Girolami.
\newblock Markov chain monte carlo from lagrangian dynamics.
\newblock {\em Journal of Computational and Graphical Statistics},
  24(2):357--378, 2015.

\bibitem{shahbaba2014split}
Babak Shahbaba, Shiwei Lan, Wesley~O Johnson, and Radford~M Neal.
\newblock Split hamiltonian monte carlo.
\newblock {\em Statistics and Computing}, 24(3):339--349, 2014.

\bibitem{ombao2016}
Hu~L. Wang~Y. and Ombao H.
\newblock {\em Statistical Analysis of Electroencephalograms}, pages 523--565.
\newblock CRC Press, 2016.

\bibitem{brockwell2013time}
Peter~J Brockwell and Richard~A Davis.
\newblock {\em Time series: theory and methods}.
\newblock Springer Science \& Business Media, 2013.

\bibitem{pennec2006riemannian}
Xavier Pennec, Pierre Fillard, and Nicholas Ayache.
\newblock A riemannian framework for tensor computing.
\newblock {\em International Journal of Computer Vision}, 66(1):41--66, 2006.

\bibitem{moakher2011riemannian}
Maher Moakher and Mourad Z{\'e}ra{\"\i}.
\newblock The riemannian geometry of the space of positive-definite matrices
  and its application to the regularization of positive-definite matrix-valued
  data.
\newblock {\em Journal of Mathematical Imaging and Vision}, 40(2):171--187,
  2011.

\bibitem{helgason1979differential}
Sigurdur Helgason.
\newblock {\em Differential geometry, Lie groups, and symmetric spaces},
  volume~80.
\newblock Academic press, 1979.

\bibitem{koh1965affine}
Sebastian~S Koh.
\newblock On affine symmetric spaces.
\newblock {\em Transactions of the American Mathematical Society},
  119(2):291--309, 1965.

\bibitem{duane1987hybrid}
Simon Duane, Anthony~D Kennedy, Brian~J Pendleton, and Duncan Roweth.
\newblock Hybrid monte carlo.
\newblock {\em Physics letters B}, 195(2):216--222, 1987.

\bibitem{nash1956imbedding}
John Nash.
\newblock The imbedding problem for riemannian manifolds.
\newblock {\em Annals of mathematics}, pages 20--63, 1956.

\bibitem{Lan2016}
Shiwei Lan and Babak Shahbaba.
\newblock {\em Sampling Constrained Probability Distributions Using Spherical
  Augmentation}, pages 25--71.
\newblock Springer International Publishing, Cham, 2016.

\bibitem{svensson2010reference}
Lennart Svensson and Magnus~Lundberg Nordenvaad.
\newblock The reference prior for complex covariance matrices with efficient
  implementation strategies.
\newblock {\em IEEE Transactions on Signal Processing}, 58(1):53--66, 2010.

\bibitem{allen2016nonspatial}
Timothy~A Allen, Daniel~M Salz, Sam McKenzie, and Norbert~J Fortin.
\newblock Nonspatial sequence coding in ca1 neurons.
\newblock {\em The Journal of Neuroscience}, 36(5):1547--1563, 2016.

\bibitem{berger2009formal}
James~O Berger, Jos{\'e}~M Bernardo, and Dongchu Sun.
\newblock The formal definition of reference priors.
\newblock {\em The Annals of Statistics}, pages 905--938, 2009.

\bibitem{sohl2014hamiltonian}
Jascha Sohl-Dickstein, Mayur Mudigonda, and Michael~R DeWeese.
\newblock Hamiltonian monte carlo without detailed balance.
\newblock {\em arXiv preprint arXiv:1409.5191}, 2014.

\bibitem{betancourt2013generalizing}
MJ~Betancourt.
\newblock Generalizing the no-u-turn sampler to riemannian manifolds.
\newblock {\em arXiv preprint arXiv:1304.1920}, 2013.

\bibitem{schwartzman2015lognormal}
Armin Schwartzman.
\newblock Lognormal distributions and geometric averages of symmetric positive
  definite matrices.
\newblock {\em International Statistical Review}, 2015.

\bibitem{fazayeli2016matrix}
Farideh Fazayeli and Arindam Banerjee.
\newblock The matrix generalized inverse gaussian distribution: Properties and
  applications.
\newblock {\em arXiv preprint arXiv:1604.03463}, 2016.

\bibitem{magnus1995matrix}
Jan~R Magnus, Heinz Neudecker, et~al.
\newblock Matrix differential calculus with applications in statistics and
  econometrics.
\newblock 1995.

\bibitem{hjorungnes2007complex}
Are Hjorungnes and David Gesbert.
\newblock Complex-valued matrix differentiation: Techniques and key results.
\newblock {\em IEEE Transactions on Signal Processing}, 55(6):2740--2746, 2007.

\end{thebibliography}

\end{document}